\documentstyle[11pt]{article}
\input psfig.tex

\newcommand{\be}{\begin{equation}}                        
 \newcommand{\ee}{\end{equation}}                         
\newcommand{\bd}{\begin{displaymath}}                     
 \newcommand{\ed}{\end{displaymath}}                      
\newcommand{\bit}{\begin{itemize}}                        
 \newcommand{\eit}{\end{itemize}}                         
\newcommand{\ben}{\begin{enumerate}}                      
 \newcommand{\een}{\end{enumerate}}                       
\newcommand{\baa}{\begin{array}{lll}}                     
 \newcommand{\eaa}{\end{array}}                           
\newcommand{\ba}{\begin{eqnarray}}                        
 \newcommand{\ea}{\end{eqnarray}}                         
\newcommand{\la}{\label}                                  
  \newcommand{\Ds}{\displaystyle}                         
\newcommand{\q}{\bar q}                                   
 \newcommand{\nn}{\nonumber}                              
  \newcommand{\xx}{\left(\bar x\rightarrow x\right)}      
\newcommand{\xxyy}                                        
 {\left(\bar x\rightarrow x,\,\bar y\rightarrow y\right)} 
\def\MSbar{\relax\ifmmode\overline                        
            {\rm MS}\else{$\overline{\rm MS}${ }}\fi}     
\def\as{\relax\ifmmode \alpha_s\else{$ \alpha_s${ }}\fi}  
\def\abar{\relax\ifmmode{\bar{a}}\else{$\bar{a}${ }}\fi}  
  \def\ie{\hbox{\it i.e.}{ }} \def\etc{\hbox{\it etc.}{ }}
   \def\eg{\hbox{\it e.g.}{ }}  
\renewcommand\thefootnote{\fnsymbol{footnote}}            
\newcounter{myfig}                                        
\newcommand{\myfig}{\refstepcounter{myfig}}               

\begin{document}
\begin{titlepage}
\vspace*{20mm}

{\flushright
    Dubna preprint, JINR E2-97-419 \\
    hep-ph/9803298 \\}

\today
\begin{center}
\vspace*{20mm}

{\Large\bf THE $\rho$-MESON AND RELATED MESON WAVE FUNCTIONS\\
            IN QCD SUM RULES WITH NONLOCAL CONDENSATES}\\[0.5cm]
\end{center}
\bigskip

\begin{center}
{\large A.~P.~Bakulev\footnote{E-mail: bakulev@thsun1.jinr.dubna.su},
  \ \ S.~V.~Mikhailov\footnote{E-mail: mikhs@thsun1.jinr.dubna.su}}\\
{\it Bogoliubov Laboratory of Theoretical Physics,\\
 Joint Institute for Nuclear Research, \\
141980, Moscow Region, Dubna, Russia}\\[0.5cm]
\end{center}
\vspace*{20mm}

\begin{abstract}
  We apply the nonlocal condensate formalism to construct
generalized sum rules (including $O(\as)$-radiative corrections)
for $\pi$-, $\rho$- and $\rho'$-meson wave functions
of twist $2$.
Besides, we predict the lepton decay constant $f_{\rho'}$
and estimate the mass of $\rho'$-meson.
For  all these mesons we obtain the first 10 moments of
longitudinal wave functions, suggest the models for them and
discuss their properties.
We consider the peculiarities of the QCD sum rules
with nonlocal condensates
for transverse $\rho$- and $b_1$-meson wave functions.
These results are compared with those found by Ball and Braun
and by Chernyak and Zhitnitsky.
\end{abstract}
\vspace {1cm}

\end{titlepage}

 \section{Introduction}
\renewcommand\thefootnote{\arabic{footnote}}
\setcounter{footnote}{0}
\la{sec-1}
An important problem in the theory of strong interactions is
to calculate hadronic wave functions $\varphi_{\pi}(x)$,
$\varphi_{\rho}(x)$, ..., $\varphi_{N}(x_1 , x_2 , x_3) $, \etc
from the first principles of QCD.
These phenomenological distributions of partons on the fraction $xP$
of a hadron momentum $P$ are a natural result
of ``factorization theorems'' applied to hard exclusive processes
\cite{cz77,ar77,fj77,bl78}.
They accumulate all the necessary information about
non-perturbative long-distance dynamics of partons in hadrons.

In the  standard QCD sum rule (SR) calculation of light meson
wave functions (WF's),
first introduced by Chernyak and Zhitnitsky (C\&Z) \cite{cz82}
and recently re-estimated by Ball and Braun (B\&B)
for the $\rho$-meson \cite{bb96},
it is assumed that the correlation length $\Lambda$
of vacuum fluctuations is large compared to
a typical hadronic scale $\sim 1/m_{\rho}$.
Thus, one can replace the original nonlocal objects
like $M(z^2)=\langle \bar q(0)E(0,z)q(z) \rangle$
\footnote{Here $E(0,z)=P\exp(i \int_0^z dt_{\mu} A^a_{\mu}(t)\tau_a)$
is the Schwinger phase factor required for gauge invariance.}
by constant quantities of $\langle\bar q(0)q(0) \rangle$-type.
Based on this hypothesis, the well-known QCD SR approach \cite{svz}
has been applied in \cite{cz82} to calculate the first two moments
$\langle\xi^N\rangle \equiv \int_{0}^{1} \varphi(x) (2x-1)^N dx$
with $N=2,~4$ for WF's of light mesons.
And just from these moments the whole WF's have been reconstructed
which are now referred to as C\&Z WF's.

Now it is known that hadronic WF's are rather sensitive
to the widths of the function $M(z^2)$ and of other nonlocal
condensates \cite{nlwf,nlwf92,br91}
and the crucial parameter $\Lambda \cdot m_\rho \sim 1$.
Therefore, one should use nonlocal condensates (NLC's) like $M(z^2)$
whose forms reflect the complicated structure of QCD vacuum.
Certainly, these objects can subsequently be expanded over
the local condensates $\langle\q(0)q(0)\rangle$,
$\langle\q(0)\nabla^2q(0)\rangle$, {\em etc.}
(here $\nabla_\mu = \partial_\mu - ig \hat{A}_{\mu}$
is the covariant derivative), and
one can come back to the standard SR by truncating this series.
Our strategy is to avoid an expansion of that sort
because we thus lose  an important physical property
of non-perturbative vacuum -- the possibility of vacuum
quarks and gluons to flow through vacuum
{\bf with non-zero virtuality} $ k^2 \neq 0$.
Indeed, the average virtuality of vacuum quarks
$\langle k^2_{q} \rangle  = \lambda_q^2$ is not small
whereas the standard approach inevitably suggests
that $k$ is not small but exact zero.
The value of $\lambda_q^2$ can be extracted from
the QCD SR analysis \cite{belioffe} and is connected
with the condensate of the next ($d=5$) dimension
\be
 \lambda_q^2
 \equiv \frac{\langle\bar q \nabla^2q\rangle}
 {\langle\bar qq\rangle}
 = \frac{\langle\bar q\left(ig\sigma_{\mu\nu}G^{\mu\nu}\right)q\rangle}
   {2\langle\bar qq\rangle}
   =  0.4 \pm 0.1 \, \mbox{GeV}^2 ,
\label{eq:lambda}
\ee
that is of an order of the typical hadronic scale,
$m_{\rho}^2 \approx 0.6$ GeV$^2$.
An estimate for $\lambda_q^2$ in the framework of the instanton liquid
model \cite{DEM97} yields a similar number,
and the lattice QCD calculations \cite{KS87} give
$\lambda_q^2 = 0.55 \pm 0.05 \, \mbox{GeV}^2$,
a similar estimate $\lambda_q^2 = 0.5 \pm 0.05 \, \mbox{GeV}^2$
has been obtained in~\cite{piv91}.

  Careful inspection of consequences of that approach to
QCD SR for pion WF \cite{nlwf,nlwf92,ar97} has revealed
that the introduction of the correlation length
$\Lambda\sim 1/\lambda_q$ into condensate distributions produces
much smaller values for the first moments of pion WF than C\&Z values.
As a result, the pion WF is strongly different in shape from
the C\&Z and approaches the asymptotic WF
$\varphi^{as}(x) \equiv 6x(1-x)$ \cite{bl78},
\ie,~$\varphi_\pi(x)\approx\varphi^{as}(x)$
\footnote{Note here other independent evidences that
the pion WF is close to its asymptotic form:
lattice calculations of the corresponding
$\langle\xi^2\rangle_{\pi}$ in \cite{dgr91};
QCD SR estimate of the magnitude of $\varphi_\pi(x)$
at the midpoint $x=1/2$ in \cite{bf89}.}
Later, this WF was confirmed by independent consideration
of the QCD SR directly for the function $\varphi_\pi(x)$ based on
the non-diagonal correlator and on the advanced smooth
distribution function for the quark nonlocal condensate
\cite{ar94,bm96}.
The key element of both the ways was to take
into account the main physical reason --
vacuum correlation length $\Lambda \sim 1/\lambda_q$
is of an order of $1/m_\rho$.

\begin{figure}[thb]
 \vspace*{-40mm}
  \[\hspace{-50mm}
   \psfig{figure=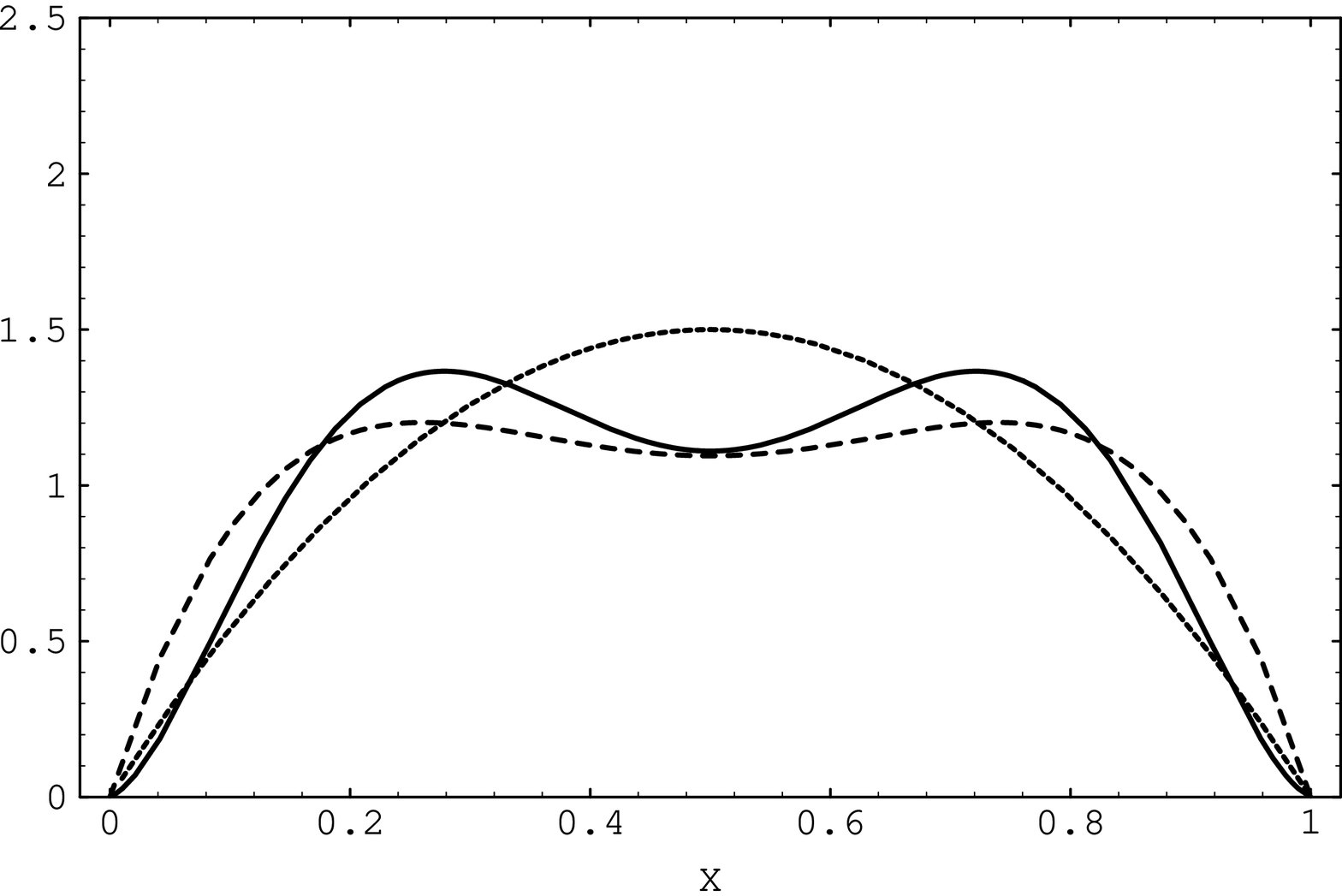,%
          bbllx=2.5cm,bblly=26cm,bburx=12.5cm,bbury=6cm,%
          width=7cm}
      \]
       \vspace{13cm}
        \caption{Longitudinal wave function of the $\rho$-meson:
                 solid line -- from NL QCD SR,
                 dotted line -- asymptotic WF, dashed line -- C\&Z WF.}
         \myfig{\label{fig:wf_lro}}
          \end{figure}

\begin{figure}[htb]
 \vspace*{-40mm}
  \[\hspace{-50mm}
   \psfig{figure=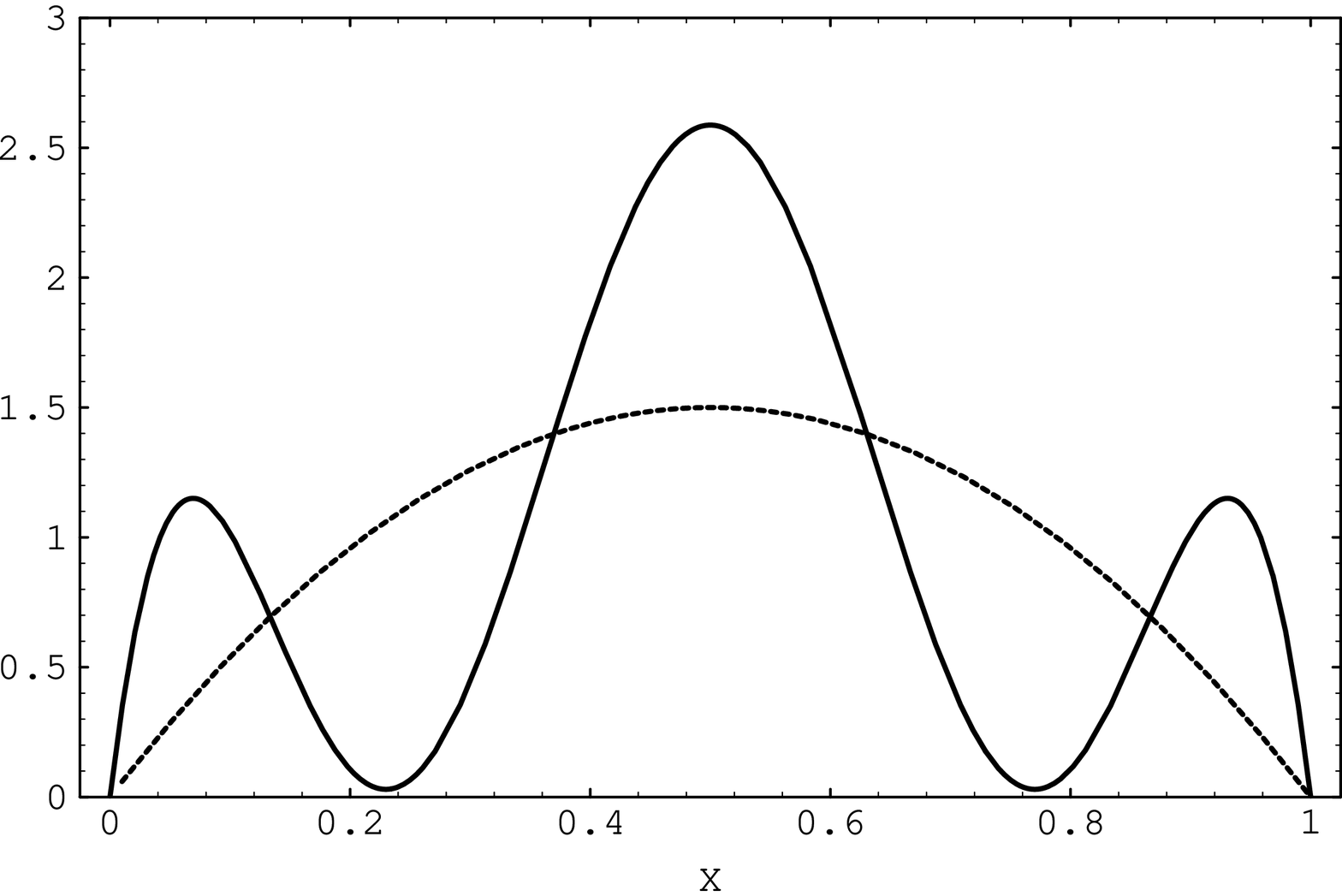,%
          bbllx=2.5cm,bblly=26cm,bburx=12.5cm,bbury=6cm,%
          width=7cm}
      \]
       \vspace{13cm}
        \caption{Longitudinal wave function of the $\rho'$-meson:
                 solid line -- from NL QCD SR,
                 dotted line -- asymptotic WF.}
          \myfig{\label{fig:wf_lrs}}
           \end{figure}

  Our goal here is to show that in the case of QCD SR
for the $\rho$-meson channel the situation is similar
to the pion one.
All the predictions of the standard QCD SR
(the C\&Z ones for the longitudinal case \cite{cz82} and B\&B
ones for the transverse case \cite{bb96} --
we call them the Local QCD SR)
for moments $\langle\xi^{N}\rangle_{\rho}$ with $N>2$
could not be considered as reliable; as we show in our analysis,
even the value of $\langle\xi^2\rangle_{\rho}$ is
indefinite.
We apply the NLC formalism to calculate
the diagonal correlators for $\rho$-meson currents,
introduced in \cite{cz82}, and construct
generalized SR including $O(\alpha_s)$-radiative corrections
to obtain WF's of twist 2. The first ten moments of longitudinal WF's
of $\rho$- and $\rho'$-mesons are estimated (Table 1) and
the models for them (see Figs.\ref{fig:wf_lro}--\ref{fig:wf_lrs})
are suggested.
Also, we predict the lepton decay constant $f_{\rho'}$
and estimate the mass of $\rho'$-meson.
We discuss the strong dependence of results
for moments of transverse WF's in $\rho$-channel
on the model of nonlocal gluon condensate contribution.
The calculation technique is the same as in Refs.
\cite{nlwf,nlwf92};
therefore, the corresponding details are omitted below,
but we shall through the text keep the connection with the
well-understood pion case.

\section{Generalized sum rules for the $\pi$- and $\rho$-meson
         channels vs the standard version}

  For the helicity-zero charged vector meson
$M_{\mid\lambda\mid=0}=M_L$, $M_L=\rho,\  \rho' \ldots $,
the leading-twist WF is defined as
\be
\langle0\mid \bar d(z)\gamma_{\mu}u(0)\mid M_L(p)\rangle
\Big|_{z^2=0}
 =  f_{M_L}^Lp_{\mu} \int^1_0 dx e^{ix(zp)}\ \varphi^L_{M_L}(x)
\ee
(In accordance with the B\&B definition,
 in this case
 $\varepsilon^{\lambda=0}_{\mu}(p)\simeq p_{\mu}/m_{M_L}$
 as $p_z \to \infty$,
 where $\varepsilon_{\mu}(p)$ is the polarization vector);
and for axial mesons ($M_A=\pi, A_1 \ldots $) as
\be
\langle0\mid \bar d(z)\gamma_\mu\gamma_5u(0)\mid M_A(p)\rangle
\Big|_{z^2=0}
 = if_{M_A}p_{\mu} \int^1_0 dx e^{ix(zp)}\ \varphi_{M_A}(x) .
\ee
The information about $\varphi^L(x)$ can be obtained from
the correlator $I_{(n0)}(q^2)$ of vector currents $V_{(n)}(y)$,
see \eg \cite{cz82}:
\be
 i\int \!dy e^{iqy}
   \langle 0|T\left\{V^+_{(0)}(y)V_{(n)}(0)\right\}|0 \rangle
 = i^n\left(zq\right)^{n+2} I_{(n0)} ,
 \, V_{(n)}(y) \equiv \bar d(y)\hat z \left(z\nabla\right)^n u(y)
\Big|_{z^2=0};
\la{cor-V}
\ee
and the information about $\varphi^A(x)$, from an analogous formula
with the substitution $V\to A$ and $A_{(n)}(y) \equiv
 \bar d(y)\hat z \gamma_5\left(z\nabla\right)^n u(y)$.
The corresponding formula for the unit helicity state,
$M_{\perp} = M_{\mid \lambda \mid=1}$,
$M_{\perp}=\rho,\  b_1 \ldots $, is as follows
\be
 \langle0\mid\bar d(z)\sigma_{\mu\nu}u(0)\mid M_{\perp}(p)\rangle
  \Big|_{z^2=0}
 = if_{M_{\perp}}^{T}
  \left(\varepsilon^{\lambda}_{\mu}(p)p_{\nu}
       -\varepsilon^{\lambda}_{\nu}(p)p_{\mu}\right)
   \int^1_0 dx e^{ix(zp)}\ \varphi^T_{M_{\perp}}(x) .
\ee
Moments of these functions $\varphi^T(x)$ are extracted
from the correlator $J_{(n0)}(q^2)$ of tensor currents
$T_{(n)}^\mu(y)$ \cite{cz82,bb96}
\be
 i\int \!dy e^{iqy}
  \langle 0|T\left\{T^{\mu+}_{(0)}(y)T^\mu_{(n)}(0)\right\}|0\rangle
 = -2i^n\left(zq\right)^{n+2} J_{(n0)} ,\,
 T^\mu_{(n)}(y)
  \equiv
   \bar d(y)\sigma^{\mu\alpha}z_\alpha\left(z\nabla\right)^n u(y) .
\la{cor-T}
\ee
This correlator $J_{(n0)}(q^2)$ contains the contribution
from states with different parity (see the analysis in \cite{bb96}).
Therefore, the contamination from $b_1$-meson
$\left(J^{PC}=1^{+-}\right)$ in the phenomenological part of SR
is mandatory.

The theoretical ``condensate'' parts for both the correlators,
$I_{(n0)}$ in (\ref{cor-V}) and $J_{(n0)}$ in (\ref{cor-T}),
contain the same 6 universal elements as for the pion case.
An exception concerning transverse sum rules (see (\ref{eq:srf_t}))
will be discussed below.
The diagram origins of these elements
$\Delta\Phi_\Gamma\left(x;M^2\right)$, where $M^2$ is the Borel
parameter, are described in detail in \cite{nlwf,nlwf92}.
By the direct SR formulation for WF, one immediately constructs
a ``daughter SR'' for any functional
of $\varphi_{M}(x)$ (not only for moments $\langle\xi^N\rangle$).
Let us write down the final SR's including WF's of $\rho$-meson and
next resonances $\rho'$ and $b_1$ into the phenomenological parts
(the corresponding SR for the pion (axial) channel is written
 below for comparison):
\ba&\Ds
 \left(f_{\pi}\right)^2\varphi_\pi(x)
 + \left(f_{A_1}\right)^2\varphi_{A_1}(x) e^{-m^2_{A_1}/M^2}
 = \int_{0}^{s_0^A}\rho^{pert}_L(x;s)e^{-s/M^2}ds +
 \phantom{.}& \label{eq:srf_p}\\&\Ds\phantom{.}
 + \Delta\Phi_G(x;M^2) + \Delta\Phi_S(x;M^2)
 + \Delta\Phi_V(x;M^2) + \Delta\Phi_{T_1}(x;M^2)
 + \Delta\Phi_{T_2}(x;M^2) + \Delta\Phi_{T_3}(x;M^2)&;\nonumber
\ea\ba&\Ds
 \left(f_\rho^L\right)^2\varphi_\rho^L(x) e^{-m^2_\rho/M^2}
 + \left(f_{\rho'}^L\right)^2\varphi_{\rho'}^L(x) e^{-m^2_{\rho'}/M^2}
 =  \int_{0}^{s_0^L}\rho^{pert}_L(x;s)e^{-s/M^2}ds +
 \phantom{.}& \label{eq:srf_l}\\&\Ds\phantom{.}
 + \Delta\Phi_G(x;M^2) - \Delta\Phi_S(x;M^2)
 + \Delta\Phi_V(x;M^2) + \Delta\Phi_{T_1}(x;M^2)
 + \Delta\Phi_{T_2}(x;M^2) + \Delta\Phi_{T_3}(x;M^2)& ;\nonumber
\ea\ba&\Ds
 \left(f_\rho^T\right)^2\varphi_\rho^T(x) e^{-m^2_\rho/M^2}
 + \left(f_{b_1}^T\right)^2\varphi_{b_1}^T(x)e^{-m^2_{b_1}/M^2}
  =  \int_{0}^{s_0^T}\rho^{pert}_T(x;s)e^{-s/M^2}ds +
 \phantom{.}& \label{eq:srf_t}\\&\Ds\phantom{.}
 + \Delta\Phi_G(x;M^2) - \Delta\Phi_G'(x;M^2)
 + \Delta\Phi_V(x;M^2) + \Delta\Phi_{T_1}(x;M^2)
 + \Delta\Phi_{T_2}(x;M^2) - \Delta\Phi_{T_3}(x;M^2)& ,\nonumber
\ea
where $s_0^{A,L,T}$ are the effective continuum thresholds
in the pion and the $\rho$-meson $L$ and $T$ cases, respectively.
Perturbative spectral densities $\rho^{pert}_{L,T}(x;s)$
are presented in an order of $O(\alpha_s)$ in \cite{nlwf,nlwf92}
for the $L$ case and in \cite{bb96} for the $T$ case
(see Appendix B).
Radiative corrections reach 10 \% of $\rho^{pert}_{L,T}(s \sim 1$ GeV$^2)$.
Contributions $\Delta\Phi_{\Gamma}(x;M^2)$ depend on a specific form
of NLC's $M(z^2)$, ..., \etc
To construct SR for WF's, it is useful to parametrize
these NLC behaviors by the ``distribution functions''
\cite{nlwf,nlwf92,ar94,bm96} {\em a'la} $\alpha$-representation
of propagators, \eg, $f_S(\nu)$ for the scalar condensate $M(z^2)$
\footnote{ In deriving these sum rules we can always make a
Wick rotation, i.e., we assume that all coordinates are
Euclidean, $z^2 <0$.}
\ba
 M\left(z^2\right)
  = \langle\bar q(0)q(0)\rangle
     \int_{0}^{\infty} e^{\nu z^2/4}\, f_S(\nu)\, d\nu,~
 \mbox{where}~\int_{0}^{\infty}\, f_S(\nu)\, d\nu = 1,\
 \int_{0}^{\infty}\nu f_S(\nu) d\nu = \frac{\lambda_q^2}2.
 \label{eq:qq}
\ea
The function $f_S(\nu)$ and other similar functions
$f_\Gamma(\nu)$ describe distributions of vacuum fields
in virtuality $\nu$ for every type of NLC.
They completely determine the r.h.s. of SR's in
(\ref{eq:srf_p})--(\ref{eq:srf_t}).
The general forms of elements $\Delta\Phi_{\Gamma}(x;M^2)$
as functionals of $f_\Gamma(\nu)$ will be published in a
separate paper. For the standard (constant) condensates
$\langle G(0)G(0)\rangle$ and $\langle q(0)q(0)\rangle$
these distributions are trivial in form, \eg,
~$f_S(\nu)=\delta(\nu)$ (Appendix A).
To include the condensates of the next ($\sim \lambda_q^2$)
and higher dimensions into consideration ,
one should add the contributions to ~$f_S(\nu)$
proportional to the derivatives of $\delta$-functions,
$\delta(\nu)^{'},~\delta(\nu)^{''}, \ldots$.
It is clear that since there is no QCD vacuum theory,
merely models of real distributions can be suggested for
these distribution functions.
However, for the purpose of QCD SR's for moments
$\langle \xi^N \rangle$ we need an approximate information
about the $f_\Gamma(\nu)$ behaviour.
Therefore, we here apply the simplest ansatz \cite{nlwf,nlwf92},
like $f_S(\nu)=\delta(\nu-\lambda_q^2/2)$,
to take into account only the main effect,
the non-zero average virtuality $\langle k^2 \rangle$ of vacuum fields.
This form of $f_S(\nu)$ can be regarded as a result
of resummation of an infinite subset
of the above-mentioned contributions
$\sim  (\lambda_q^2/2)^n~ \delta(\nu)^{\{n\}}$ connected with the
single scale $\lambda_q^2$ \cite{nlwf,nlwf92}.
The corresponding expressions for $\Delta\Phi_{G,S,V,T_i}(x;M^2)$
are collected in Appendix A.

  Now let us take the limits $\lambda_q^2 \to 0$,
~$\Delta \Phi_{\Gamma}(x, M^2)\to\Delta \varphi_{\Gamma}(x, M^2)$
and $\rho^{pert}_{L}(x,s) \to \rho^{Born}_{L}(x,s)$ for SR
in eq.(\ref{eq:srf_l}) to return to the standard
approach (see these reduced elements in Appendix A).
We try to inspect the subtle points and the range of validity
of C\&Z SR.
These authors extracted $\langle\xi^{N}\rangle$
exactly in the same way as the $f_{\rho}$ value
(B\&B limited themselves only to extraction of
 $\langle\xi^2\rangle$).
However, the nonperturbative terms in their sum rule
(the $\rho'$-contribution is omitted for simplicity)
have a completely different $N$-dependence compared to
the perturbative one and, {\em a priori,} it is not clear
whether a straightforward use of the `$N=0$ technology'
can be justified for higher $N$
(for definiteness, we consider here only the $\rho$-meson
 (longitudinal) case;
 the same arguments apply also to the pion case, see criticism
 in \cite{nlwf92,br91,ar97})
\ba&&
\left(f_\rho^L\right)^2\langle\xi^N\rangle_\rho^L e^{-m^2_\rho/M^2}
 + \frac{3M^2}{4\pi^2(N+1)(N+3)}e^{-s_N/M^2} = \nonumber \\ &&
 =\ \frac{3M^2}{4\pi^2(N+1)(N+3)}
 +  \frac{\langle(\alpha_s/\pi)GG\rangle}{12 M^2}
 + \frac{16}{81}\pi(4N-7)
    \frac{\langle\sqrt{\alpha_s}\bar qq\rangle^2}{M^4}.
\la{eq:czsr}
\ea
The scale determining the magnitude of all hadronic parameters
including $s_N$ (the ``continuum threshold'' \cite{svz})
is eventually settled by the ratios of condensate contributions
to the perturbative term.
If the condensate contributions in the C\&Z sum rule (\ref{eq:czsr})
had the same $N$-behavior as the perturbative term,
the $N$-dependence of $\langle\xi^N\rangle$
would be determined by the overall factor $3/(N+1)(N+3)$
and the resulting WF $\varphi (x)$ would coincide
with the ``asymptotic'' form $\varphi^{as}(x)$.

  However, the ratios of the $\langle\bar qq\rangle$- and
$\langle GG\rangle$-corrections to the perturbative term
in eq.(\ref{eq:czsr}) are growing functions of $N$.
This reduces the predictable power of the local QCD SR's
with the growth of $N$.
To reveal consequences of this effect more clearly,
let us consider the so-called SR fidelity windows,
\i.e., regions of the Borel parameter $M^2$ where one should
obtain reliable SR predictions.
In accordance with the QCD SR practice \cite{svz},
these fidelity windows are determined by two conditions:
the lower bound $M^2_-$ by demanding that the relative value
of $\langle GG\rangle$- and  $\langle\bar qq\rangle$-contributions
to OPE series should not be larger than 30\%,
the upper one $M^2_+$ by requiring that a relative contribution
of higher states in the phenomenological part of SR should not be
larger than 30\%.
Suggesting independence of the threshold of $N$
($s_N \approx s_0 \approx 1.5$~GeV$^2$), we have in the case of $N=0$:
$M^2_-=0.4$~GeV$^2$, $M^2_+=1.34$~GeV$^2$.
But for $N=2$ we have $M^2_-=0.73$~GeV$^2$, $M^2_+=1.34$~GeV$^2$,
and for $N=4$ -- even $M^2_-=1.5$~GeV$^2$, $M^2_+=1.34$~GeV$^2$.
That is, the fidelity window shrinks to an empty set in the last case.
C\&Z had to suggest that $s_2 \approx 1.9$~GeV$^2$ and
$s_4 \approx 2.2$~GeV$^2$ to extend the fidelity windows and
to obtain any stability with respect to $M^2$.
It is difficult to imagine such a strange type of a spectral model,
but there are no principal objections.

The situation with $\langle\xi^2\rangle_{\rho}$ is
to a certain extent special: though there exists a non-empty
fidelity window
of Local QCD SR for this quantity,
the stability of SR in this window is rather poor
and this is due to overestimation of
vacuum condensate contributions.
To improve the situation, C\&Z suggest
that the continuum threshold should be increased,
which means assigning a part of higher state contributions
to the ground state contribution, that is,
overestimation of $\langle\xi^2\rangle_{\rho}$.
On the contrary, the authors of \cite{bb96} put
the threshold unchanged, $s_2=s_0$,
and give the prediction without establishing any stability
as a mean value of SR at $M^2=0.8$ GeV$^2$ and at $M^2=2.0$ GeV$^2$;
the overestimation of $\langle\xi^2\rangle_{\rho}$
in this case is due to that of the condensate part of SR.

  In our opinion, there is no need to propose such
an exotic spectral model ($s_N=s_0+\mbox{const}\cdot N$)
because the reason for this ``exploding'' behaviour of Local SR
is quite evident, namely, a completely different dependence
on $N$ of the perturbative (the first term in the second line
of Eq.(\ref{eq:czsr})) contribution and of condensate ones.
And the origin of this difference
was explained in a series of papers
\cite{nlwf92,br91,ar97}; this is due to the
Taylor expansion of initial nonlocal objects like
$\langle\bar q(0)E(0,z)q(z)\rangle$ in powers of $z$.
The first constant term of this expansion,
$\langle\bar qq\rangle$, produces an $(N)^0$-dependent term
in SR (\ref{eq:czsr}); the next term, an $(N)^1$-dependent, and
so on.

  On the contrary, the NLC terms
$\Delta\Phi_{\Gamma}(x;M^2)$ in (\ref{eq:srf_l})
and (\ref{eq:srf_t}) lead to the moments
$\langle \xi^N\rangle$ which well decay with $N$ growing;
so a physically motivated $N$-independent continuum
threshold $s_0^{L}$ naturally appears in the SR processing.

  \section{The moments and the models of wave functions}
\begin{figure}[thb]
 \vspace*{-40mm}
  \[\hspace{-50mm}
   \psfig{figure=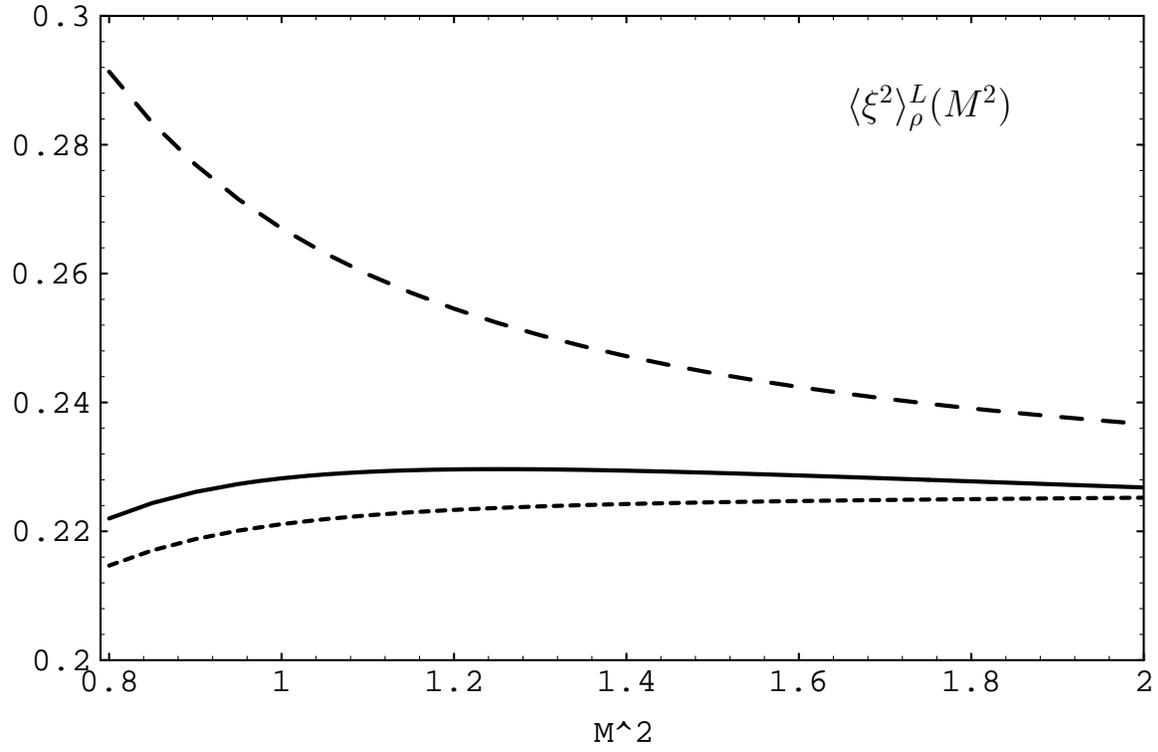,%
          bbllx=2.5cm,bblly=26cm,bburx=12.5cm,bbury=6cm,%
          width=7cm}
      \]
       \vspace{13cm}
        \caption{Curves $\langle \xi^2 \rangle_\rho^L$ as functions
                 of Borel parameter $M^2$:
                 solid line -- from NLC SR (\protect\ref{eq:srf_l})
                 with $s_0=2.4$~GeV$^2$,
                 short-dashed line -- from NLC SR without
                 $\rho'$-meson contribution ($s_0=1.5$~GeV$^2$),
                 dashed line -- from local SR of B\&B\protect\cite{bb96}
                 ($s_0=1.5$~GeV$^2$).}
         \myfig{\label{fig:x2_lr}}
          \end{figure}

  Before analyzing the results of processing of SR (\ref{eq:srf_l})
and (\ref{eq:srf_t}) for the moments $\langle \xi^N \rangle_M$,
let us consider the peculiarity of the QCD SR structure
represented in the RHS's of Eqs. (\ref{eq:srf_p}-\ref{eq:srf_t}).
Unlike the $\pi$-meson case, the contribution of the numerically
most significant ``four-quark condensate''
$\Delta\Phi_S\left(x,M^2\right)$ \cite{nlwf} is equal to zero
(for the $T$ case, see also the sign of $\Delta\Phi_{T3}$)
or even has the opposite sign (for the $L$ case).
For this reason, the role of the vacuum interaction
for the $\rho $-meson is weaker than for the pion.
As a consequence of such an SR structure the nonlocal effects
partially compensate themselves.
Therefore, the extracted values of $\langle \xi^2 \rangle_M$
in the framework of NLC SR don't drastically differ
from the results of B\&B obtained in the standard way
\footnote{Note here that the results of the $T$ case in an original
CZ work should not be taken as a pattern for a standard SR,
there is an error in the sign of the quark condensate contribution,
see \cite{rry85,bb96}}.
However, the sensitivity and stability of NLC SR are much better
than for the standard one,
compare the accuracy for the first moments in Table~1.
This allows us to estimate the first ten moments in the L case
for $\rho$- and $\rho'$-mesons.
For clarity see Fig.\ref{fig:x2_lr}, where the curves for
$\langle \xi^2 \rangle_{\rho}^L$ are represented as functions
of the Borel parameter $M^2$ in the range of stability window
for two types of QCD SR's:
  the solid line is determined from NLC SR (\ref{eq:srf_l});
  and the dashed one, from B\&B SR~\cite{bb96}
(actually B\&B paper contains only the graphics
 for the coefficient $a_2^{L}$,
 but it is straightforward to obtain the corresponding curve
 for $\langle \xi^2 \rangle_{\rho}^L = 0.2+\frac{12}{35}a_2^L$;
 for comparison we also depict in Fig.\ref{fig:x2_lr}
 the short-dashed curve for the NLC SR prediction without
 $\rho'$-meson contribution,
 because B\&B SR has been processed just in this manner).
As one can see, in the local SR case there is no stability at all.

  We have determined the following values for practically
$N$-independent continuum thresholds: $s_0^{L} \approx 2.4$ GeV$^2$.
Fidelity windows for the $L$ case are:
$0.6~\mbox{GeV}^2 \le M^2 \le 2.0~\mbox{GeV}^2$
for all $N=0, 2, \ldots, 10$.
The ranges of stability within these fidelity windows almost
coincide with these windows, starting for all $N$
at $M^2\approx 0.8$ GeV$^2$.
\\ \vspace{2mm}

\noindent\hspace*{0.045\textwidth}
\begin{minipage}{0.91\textwidth}
\begin{tabular}{|cl||c|c|c|c|c|c|}\hline
&&\multicolumn{6}{|c|}{} \\
&&\multicolumn{6}{|c|}{The moments $\langle \xi^N \rangle_M(\mu^2)$
at $\mu^2 \sim 1$ GeV$^2$}\\
&&\multicolumn{6}{|c|}{
(errors are depicted in brackets following a standard manner)}\\
&&\multicolumn{6}{|c|}{} \\ \cline{3-8}
 & & & & & & &\\
\multicolumn{2}{|c||}{Type of SR}
   & $f_{M}\left(\mbox{GeV}^2\right)$
     &$\hspace{0.1mm}N=2\hspace{0.1mm}$
       &$\hspace{0.1mm}N=4\hspace{0.1mm}$
         &$\hspace{0.1mm}N=6\hspace{0.1mm}$
           &$\hspace{0.1mm}N=8\hspace{0.1mm}$
             &$\hspace{0.1mm}N=10\hspace{0.1mm}$ \\
 & & & & & & & \\ \hline \hline
\multicolumn{2}{|c||}{\strut\vphantom{\vbox to 6mm{}}
  Asympt. WF$_{\vphantom{\vbox to 4mm{}}}$}
    & $1$ & $0.2$  & $0.086$
        & $0.047$ & $0.030$ & $0.021$
   \\  \hline \hline
 {\strut\vphantom{\vbox to 6mm{}} NLC
  $_{\vphantom{\vbox to 4mm{}}}$:}
 &$\pi$
   & $0.131(2)$ & $0.25(1)$ & $0.110(7)$
         & $0.054(3)$ & $0.031(2)$ & 0.0217(7)
    \\ \cline{3-8}
 {\strut\vphantom{\vbox to 6mm{}} C\&Z
  $_{\vphantom{\vbox to 4mm{}}}$:}
 & $\pi$
   & $0.131$    & $0.40$    & $0.24$
         & --         & --         & --
    \\ \hline \hline
 {\strut\vphantom{\vbox to 6mm{}} NLC
  $_{\vphantom{\vbox to 4mm{}}}$:}
 &$\rho^L$
   & $0.201(5)$ & $0.227(7)$ & $0.095(5)$
         & $0.051(4)$ & $0.030(2)$ & $0.020(5)$
    \\ \cline{3-8}
 {\strut\vphantom{\vbox to 6mm{}} B\&B
  $_{\vphantom{\vbox to 4mm{}}}$:}
 & $\rho^L$ & $0.205(10)$ & $0.26(4)$ & -- & -- & -- & --
    \\ \cline{3-8}
 {\strut\vphantom{\vbox to 6mm{}} C\&Z
  $_{\vphantom{\vbox to 4mm{}}}$:}
 & $\rho^L$ & 0.194 & 0.26 & 0.15 & -- & -- & --
    \\ \hline \hline
 {\strut\vphantom{\vbox to 6mm{}} NLC
  $_{\vphantom{\vbox to 4mm{}}}$:}
 & $\rho'^L$
    & $0.175(10)$ & $0.226(10)$  & $0.145(7)$
        & $0.106(5)$ & $0.082(4)$ & $0.064(4)$
   \\  \hline \hline
 {\strut\vphantom{\vbox to 6mm{}} NLC
  $_{\vphantom{\vbox to 4mm{}}}$:}
 & $\rho^T$
   & $0.169(5)$ & $0.325(10)$  & ?
         & ? & ? & ?
    \\ \cline{3-8}
 {\strut\vphantom{\vbox to 6mm{}} B\&B
  $_{\vphantom{\vbox to 4mm{}}}$:}
 & $\rho^T$
   & $0.160(10)$ & $0.27(4)$ & -- & -- & -- & --
    \\ \cline{3-8}
 {\strut\vphantom{\vbox to 6mm{}} C\&Z
  $_{\vphantom{\vbox to 4mm{}}}$:}
 & $\rho^T$
   & 0.200 & 0.15 & $\le 0.06$ & -- & -- & -- \\ \hline \hline
 {\strut\vphantom{\vbox to 6mm{}} NLC
  $_{\vphantom{\vbox to 4mm{}}}$:}
 & $b_1^T$
   & $0.181(5)$ & $0.090(5)$ & ?
         & ? & ? & ?
    \\ \cline{3-8}
 {\strut\vphantom{\vbox to 6mm{}} B\&B
  $_{\vphantom{\vbox to 4mm{}}}$:}
 & $b_1^T$
    & 0.180-0.170 & -- & -- & -- & -- & --
    \\ \hline \hline
\end{tabular}\\[2.5mm]

\noindent{\bf Table 1}.
The moments $\langle \xi^N \rangle_M(\mu^2)$ at
$\mu^2 \sim 1$ GeV$^2$. C\&Z give all moments normalized
to the normalization point $\mu_0=500$ MeV;
here we present these moments normalized
to the normalization point $\mu=1$ GeV.
\end{minipage}
\vspace{1.8mm}

For transverse SR (\ref{eq:srf_t}),
in its theoretical side, we face the problem
that an important part of nonlocal gluon contribution
$\Delta\Phi'_{G}\left(x,M^2\right)$ is not yet
estimated.
We have suggested a naive model, instead of the local result
$\Ds \Delta\varphi'_{G}\left(x,M^2\right)
 \equiv \frac{\langle \alpha_s GG \rangle}{6\pi M^2}$
we put
$$ \Delta\Phi'_{G}\left(x,M^2\right) =
    \Delta\varphi'_{G}\left(x,M^2\right)
     \frac{\theta\left(\Delta<x\right)\theta\left(x<1-\Delta\right)}
           {1-2\Delta}.$$
This simulation eliminates end-point ($x=0, 1$) effects
due to the influence of the vacuum gluon nonlocality,
which is inspired by the analysis in \cite{ms93}
and our experience in the nonlocal quark case (see Appendix A).
For the zero-th moment SR ($N=0$) any model is not too important,
they give a gluonic contribution close to one
in the local SR, and we obtain the best stability
with $s_0^{T} \approx 2.3$ GeV$^2$ and
for $0.75~\mbox{GeV}^2 \le M^2 \le 1.9~\mbox{GeV}^2$.
But for the next moment the results are very sensitive
to the model of gluon contribution.
Indeed, these results (see Table $1$) drastically change
if we leave the local expression
$\Ds \Delta\varphi'_{G}\left(x,M^2\right)$ unchanged:
 $\langle \xi^2 \rangle_\rho^T = 0.231(8)$ and
 $\langle \xi^2 \rangle_{b_1}^T = 0.220(8)$).
So, we realize that our modelling is of great importance for
the results discussed and we can't give a reliable prediction for
 $\langle \xi^{2N} \rangle_{\rho,b_1}^T$ ($N = 1, \ldots $)
without essential calculational efforts.

We also re-estimate the first ten moments for the pion case
and confirm the main previous results
of nonlocal approach \cite{nlwf92}.
In this axial channel (see (\ref{eq:srf_p})) we include the pion and
$A_1$-meson into the phenomenological part of SR and obtain:
$s_0^{A} \approx 2.3$ GeV$^2$, and the ranges of stability are
$0.75~\mbox{GeV}^2 \le M^2 \le 1.9~\mbox{GeV}^2$.

\begin{figure}[thb]
 \vspace*{-40mm}
  \[\hspace{-50mm}
   \psfig{figure=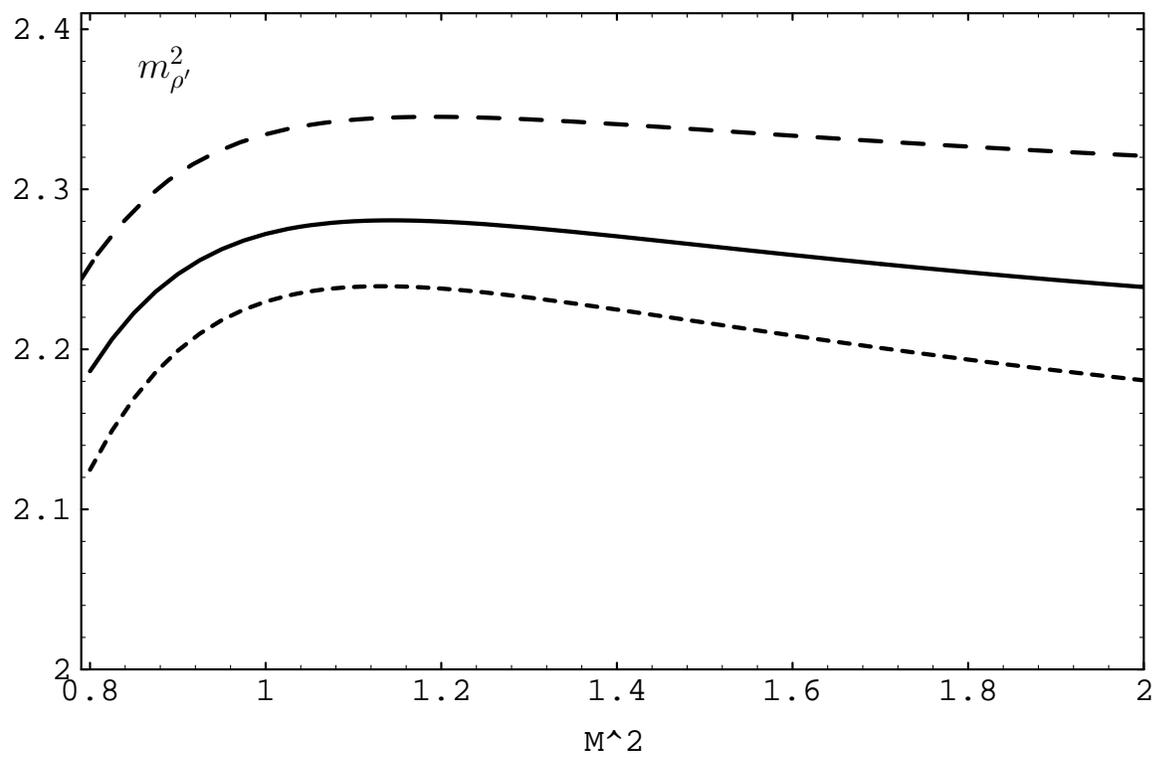,%
          bbllx=2.5cm,bblly=26cm,bburx=12.5cm,bbury=6cm,%
          width=7cm}
      \]
       \vspace{13cm}
        \caption{Extracted squared mass of the $\rho'$-meson (in GeV$^2$):
                 dashed line -- $s_0=2.9$~GeV$^2$,
                 solid line -- $s_0=2.6$~GeV$^2$,
                 dotted line -- $s_0=2.3$~GeV$^2$.}
         \myfig{\label{fig:m_rs}}
          \end{figure}

Another evidence of the efficiency of NLC SR
is the estimate of the $\rho'$-meson mass.
First, the $\rho'$-resonance with tabular mass $m_{\rho'}=1465$~MeV
was inserted into SR (\ref{eq:srf_l}) to improve the stability.
But at the second step $m_{\rho'}$ was estimated from our SR,
see Fig.\ref{fig:m_rs}.
It appears to be rather close to the experimental value \cite{r-m1}
\be
 m_{\rho'}^{theor} = 1496\pm37\ \mbox{MeV}\ , \qquad
 m_{\rho'}^{exp} = 1465\pm22\ \mbox{MeV}.
\ee
Possible models of WF's corresponding to the moments in Table 1
have the form
\footnote{Another model for $\varphi_{\pi}$,
$\varphi_{\pi}^{mod1}=\varphi^{as}(x)(1+\frac{4}{27}C^{3/2}_2(\xi))$,
corresponding to the first three moments, has been suggested in
~\cite{nlwf92}.}
\ba
 \varphi_\pi^{mod}(x, \mu^2) &=&\varphi^{as}(x)\,
     6.739\, \left(\varphi^{as}(x)\right)^{1.305}\,
      \left(x^2 + \left(1-x\right)^2\right)^{4.01} ,
\label{eq:mod_pi} \\
 \varphi_\rho^{L,mod}(x, \mu^2) &=& \varphi^{as}(x)
 \left(1+0.077 \cdot C^{3/2}_2(\xi)
        -0.077 \cdot C^{3/2}_4(\xi) \right) ,
\label{eq:mod_lro} \\
 \varphi_{\rho'}^{L,mod}(x, \mu^2) &=& \varphi^{as}(x)
 \left(1+0.075 \cdot C^{3/2}_2(\xi)
        +0.4   \cdot C^{3/2}_4(\xi)
        -0.04  \cdot C^{3/2}_6(\xi)\right) ,
\label{eq:mod_lrs}
\ea
where $\xi\equiv 1-2x$, $C^{\nu}_{n}(\xi)$ are the
Gegenbauer polynomials and $\mu^2\simeq 1$ GeV$^2$
corresponds to an average value of $M^2$.
The important problem of sensitivity of the results
to the condensate parameters in SR's will be
addressed to in a separate paper.
Here we only note that the main qualitative conclusions
concerning the shapes of $\varphi_{\rho, \rho'}^{L}$ and
$\varphi_{\pi}^A$ do not change under accessible variation of
$\lambda^2_q$.

To check the reliability of these models,
let us estimate the functional
$\Ds I[\varphi_M]=\int^1_0 \frac{\varphi_M(x)}{x}dx$
that is often used in calculations of different form factors.
This integral naturally appears in the perturbative QCD approach
to the electromagnetic form factor $F_\pi$ \cite{cz77,ar77},
to the evaluation of $\gamma\gamma^{*}\to\pi^0$ \cite{rr96}
and also to the QCD description of heavy meson
semileptonic decays \cite{bb97}.
 It is clear that $I[\varphi_M]$ is a new independent
(of moments $\langle\xi^N\rangle_M$) quantity.
Besides, the values of $I[\varphi_M]$ allow us to better
discriminate between different models for the same $\varphi_M(x)$.
The $I[\varphi_M]$ can be obtained in two different ways,
\bit
\item from QCD SR adapted to the quantity $I[\varphi^{L}]$
      by integration with weight $1/x$;
\item by direct integration of the WF models
      (\ref{eq:mod_pi})--(\ref{eq:mod_lrs}).
\eit
 Let us first consider the results for the pion case.
The range of stability here is as wide as
$0.6~\mbox{GeV}^2 \le M^2 \le 1.9~\mbox{GeV}^2$
and the quality of stability is very high.
As it is seen from Table 2 (columns 3 and 4), our estimates for
$I[\varphi_\pi]$, obtained in both the mentioned ways,
are consistent with each other within 10\% accuracy.
Moreover, the estimation for $I[\varphi_\pi]$ from SR,
is quite close to the one obtained independently
by using the nondiagonal NLC QCD SR in \cite{bm96} (column 5).
So, we may conclude that the nonlocal approach leads
to self-consistent results which are not far from the estimate
of $I[\varphi_\pi] \approx 2.4$ in \cite{rr96}
\footnote{ The authors of \cite{rr96} state that the value
$I\approx 2.4$ may be underestimated not more than by $20\%$}
(see column 6) and can be reliably discriminated from the C\&Z ones.

 A similar situation holds also for the $\rho_L$-results,
the stability is also of high quality in
the range $0.5~\mbox{GeV}^2 \le M^2 \le 2.0~\mbox{GeV}^2$.
Agreement of both kinds of the results for $I[\varphi^L]$
is rather good for the $\rho_L$-case
(the discrepancy is smaller than $5\%$)
and worse for the $\rho^{'}_L$-case.
The models (\ref{eq:mod_pi})--(\ref{eq:mod_lro})
confirm the property of NLC SR's that WF's
of a meson in the ``ground state"approaches
the  asymptotic WF (due to nonlocality effects),
proposed by Radyushkin \cite{ar97}.
The curve $\varphi_{\rho,mod}^{L}(x)$ (see Fig.\ref{fig:wf_lro})
is not far from the asymptotic WF curve and
looks close to the naive B\&B model.
Note, nevertheless, that the latter visual closeness seems
rather crude and does not allow
quantitative conclusions, \eg, NLC SR provides the values
of moments $\langle\xi^2\rangle_{\rho}$, $I[\varphi^L]$
in $15\%$ smaller than in the B\&B case.
One may expect that WF's of resonances would oscillate by analogy
with the pion resonance $\pi'$ case \cite{ar94,bm96}.
Indeed, the shape of WF $\varphi_{\rho'}^{L,mod}(x)$
(see Fig.\ref{fig:wf_lrs})
for resonance looks similar to WF of $\pi'$.
\vspace*{2mm}

\noindent\hspace*{0.08\textwidth}
\begin{minipage}{0.84\textwidth}
\begin{tabular}{|c|c|c|c|c|c|c|c|}\hline
 & & & & & & & \\
 &\hspace{0.1mm}WF\hspace{0.1mm}
   &\hspace{0.1mm}SR\hspace{0.1mm}
     &\hspace{0.1mm}WF\hspace{0.1mm}
       &\hspace{0.1mm}WF\hspace{0.1mm}
         &\hspace{0.1mm}SR\hspace{0.1mm}
           &\hspace{0.1mm}WF\hspace{0.1mm}
             &\hspace{0.1mm}WF\hspace*{0.1mm}\\
 &\hspace{0.1mm}(asymp)\hspace{0.1mm}
   &\hspace{0.1mm}(here)\hspace{0.1mm}
     &\hspace{0.1mm}(here)\hspace{0.1mm}
       &\hspace{0.1mm}(\cite{bm96})\hspace{0.1mm}
         &\hspace{0.1mm}(\cite{rr96})\hspace{0.1mm}
           &\hspace{0.1mm}(B\&B)\hspace{0.1mm}
             &\hspace{0.1mm}(C\&Z)\hspace*{0.1mm}\\
 & & & & & & & \\ \hline
 & & & & & & & \\
     $\hspace{3mm}I[\varphi_\pi]\hspace{3mm}$
 &$3$
   &$2.75\pm0.05$
     &$3.03$
       &$2.8$
         &$2.4 (\le 2.9)$
           & ---
             &$5.0$  \\ & & & & & & & \\ \hline
 & & & & & & & \\
$\hspace{3mm}I[\varphi_\rho^L]\hspace{3mm}$
 &$3$
   &$3.1 \pm 0.1$
     &$3.0$
       & ---
         & ---
           &$3.54$
             &$4.38$ \\ & & & & & & & \\ \hline
 & & & & & & & \\
$\hspace{3mm}I[\varphi_{\rho'}^L]\hspace{3mm}$
 &$3$
   &$4.7 \pm 0.2$
     &$4.3$
       & ---
         & ---
           & ---
             & ---   \\ & & & & & & & \\ \hline
\end{tabular}\\[2.5mm]

\noindent{\bf Table 2}. On the analysis of SR self-consistency,
based on evaluations of the functional
$\Ds I[\varphi_M]=\int^1_0 \frac{\varphi_M(x)}{x}dx$
in different approaches.
\end{minipage}
\vspace*{1.8mm}

 \section{Conclusion}
  \la{sect-8}
  Our basic interest in the present paper is to explore
the well-working method of NLC QCD SR in analyzing
WF's in the $\rho$-meson vector and tensor channels.
As has been noted in the previous papers
\cite{nlwf,nlwf92,br91,ar97,ar94,bm96}, just in problems
of ``nonlocal" characteristics of hadrons, such as wave functions,
form factors, \etc,
one should use the formalism of nonlocal condensates.
Let us summarize the main results of this paper:
\begin{enumerate}
\item The generalized sum rules for WF's of the $\rho$-meson and
 related resonances with nonlocal condensates are constructed.
Using the simplest ansatz \cite{nlwf92} for nonlocal quark
 condensates we obtain new estimates for the first ten moments
 of the $\rho$-meson and its resonance WF's.
 It should be emphasized that analogous evaluation within
the standard QCD SR approach is impossible.
\item We suggest the models for (see Figs.$3$--$4$)
 longitudinal WF's of $\pi$-, $\rho$- and $\rho'$-mesons
 and verify their self-consistency.
 The form of the obtained longitudinal $\rho$-meson WF is not
 far from the asymptotic WF (this conclusion noticeably differs
 from the results of the Local QCD SR \cite{cz82}).
\item As a by-product, we predict the lepton decay constant
 $f_{\rho'}=0.175 \pm 0.005$~GeV$^2$
 and estimate the mass of the $\rho'$-meson,
 $m_{\rho'}=1496 \pm 37$~MeV,
  which is now under experimental investigation \cite{r-m1}.
\item For $\pi$- and $\rho^L$-channels we perform processing of
 SR directly for the integral
 $\Ds I[\varphi]=\int^1_0 \frac{\varphi_M(x)}x dx$
 and obtain high-stable predictions for these quantities.
 Their values unambiguously favor WF's close to the asymptotic form
 rather than to the C\&Z-one.
 Moreover, the result for $\pi$-meson is also consistent
 with the model WF obtained in a special variant of NLC QCD SR's
 (in the non-diagonal QCD SR approach \cite{ar94,bm96}
 which directly connects $\pi$-WF only with nonlocal condensates)
 and with analysis of the $\gamma\gamma^{*}\to\pi^0$
 transition form factor \cite{rr96}.
\end{enumerate}
 So we can conclude that the nonlocal QCD SR approach
is self-consistent and gives the results of high stability.
To complete the calculation for transverse $\rho$-meson WF,
we need an exact result for the nonlocal gluon contribution to SR.
An open problem of this approach is the determination
of well-established models of distribution functions
$f_\Gamma(\nu)$ from the theory of nonperturbative QCD vacuum.

\vspace*{5mm}

{\large \bf Acknowledgments}
\vspace*{3mm}

{We are grateful to A.~V.~Radyushkin who inspired this
work many years ago. We are also grateful to  H.~G.~Dosch,
A.~Kataev, N.~Kochelev, R.~Ruskov and N.~Stefanis
for fruitful discussions of the main results.
A special gratitude we express to A.~Dorokhov and R.~Ruskov for careful
reading of the manuscript and giving many useful notes.
We are indebted the Russian Foundation for Fundamental Research
(contract 96-02-17631) and the Heisenberg--Landau Program for
financial support.
We are grateful to Prof. H.~G.~Dosch for
warm hospitality at the Institute for Theoretical Physics
at Heidelberg University
and to Prof. K.~Goeke and N.~Stefanis
for warm hospitality at Bochum University.}

\begin{appendix}
\vspace*{12mm}
\appendix
\hspace*{2mm}{\Large \bf Appendix}
\vspace{-3mm}
\section{Expressions for nonlocal contributions to SR}
\renewcommand{\theequation}{\thesection.\arabic{equation}}
\la{subs-A.1}
 \setcounter{equation}{0}
For vacuum distribution functions $f_{\Gamma}(\nu)$ we use
the set of the simplest ansatzes
\ba
 f_S(\nu) &=& \delta\left(\nu-\lambda_q^2/2\right)\ ;\qquad
 f_V(\nu)\ =\ \delta^\prime\left(\nu-\lambda_q^2/2\right)\ ;
\la{eq:ansv}\\
 f_{T_i}(\alpha_1,\alpha_2,\alpha_3)
          &=& \delta\left(\alpha_1-\lambda_q^2/2\right)
               \delta\left(\alpha_2-\lambda_q^2/2\right)
                \delta\left(\alpha_3-\lambda_q^2/2\right)\ .
\label{eq:anstril}
\ea
Their meaning and relation to initial NLC's
have been discussed in detail in \cite{nlwf,nlwf92}
The contributions of NLC's
$\Delta \Phi_{\Gamma}(x, M^2)$ corresponding to these ansatzes
are shown below;
the limit of these expressions to the standard (local) contributions
$\varphi_{\Gamma}(x, M^2)$ -- $\lambda_q^2\to0$,
$\Delta \Phi_{\Gamma}(x, M^2)\to\Delta \varphi_{\Gamma}(x, M^2)$
are also written for comparison.
Here and in what follows $\Delta \equiv \lambda_q^2/(2M^2)$,
$\bar\Delta\equiv 1-\Delta$:
\ba
 \Delta\Phi_S\left(x,M^2\right)
  &=&\frac{A_S}{M^4}
      \frac{18}{\bar\Delta\Delta^2}
       \left\{
        \theta\left(\bar x>\Delta>x\right)
         \bar x\left[x+(\Delta-x)\ln\left(\bar x\right)\right]
       + \xx + \right. \nonumber \\
&&\qquad\qquad
\left. + \theta(1>\Delta)\theta\left(\Delta>x>\bar\Delta\right)
         \left[\bar\Delta
              +\left(\Delta-2\bar xx\right)\ln(\Delta)\right]
        \right\},
\la{eq:phi_s}\\
 \Delta\varphi_S\left(x, M^2\right)
  &=&\frac{A_S}{M^4}9
      \left(\delta(x)+\xx\right); \nn \\
 \Delta\Phi_V\left(x, M^2\right) &=& \frac{A_S}{M^4}
     \left(x\delta'\left(\bar x-\Delta\right)+\xx\right), \\
 \Delta\varphi_V\left(x, M^2\right) &=& \frac{A_S}{M^4}
     \left(x\delta'\left(\bar x\right)+\xx\right); \la{eq:phi_v}\\
 \Delta\Phi_{T_1}\left(x,M^2\right)
  &=& -\frac{3 A_S}{M^4}\left\{
      \left[\delta(x-2\Delta) - \delta(x-\Delta)\right]
       \left(\frac1{\Delta} - 2\right)\theta(1>2\Delta)
       + \theta(2\Delta>x) \cdot\nn \right.\\ &&\left.
    \theta(x>\Delta)\theta(x>3\Delta-1)
  \frac{\bar x}{\bar\Delta}
  \left[\frac{3x}{\Delta} - 6 - \frac{1+\bar x}{\bar\Delta}\right]
  \right\} + \xx,
\la{eq:phi_t1} \\
 \Delta\varphi_{T_1}\left(x, M^2\right) &=& \frac{3A_S}{M^4}
     \left(\delta'\left(\bar x\right)+\xx\right); \nn \\
 \Delta\Phi_{T_2}\left(x,M^2\right)
  &=& \frac{4 A_S}{M^4}\bar x
     \left\{\frac{\delta(x-2\Delta)}{\Delta}\theta(1>2\Delta)
           -\theta(2\Delta>x)\theta(x>\Delta)\theta(x>3\Delta-1) \cdot
         \right. \nonumber \\ &&\left.
             \frac{1+2x-4\Delta}{\bar\Delta\Delta^2}\right\}
  + \xx,
\la{eq:phi_t2}\\
 \Delta\varphi_{T_2}\left(x, M^2\right) &=& -\frac{2 A_S}{M^4}
    \left(x\delta'\left(\bar x\right)+\xx\right); \nn \\
 \Delta\Phi_{T_3}\left(x, M^2\right)
  &=& \frac{3 A_S\bar x}{M^4\bar\Delta\Delta}
     \left\{\theta(2\Delta>x)\theta(x>\Delta)\theta(x>3\Delta-1)
      \left[2-\frac{\bar x}{\bar\Delta}-\frac{\Delta}{\bar\Delta}
      \right]\right\} \nn \\
      & & + \xx, \\
 \Delta\varphi_{T_3}\left(x, M^2\right) &=& \frac{3 A_S}{M^4}
     \left(\delta\left(\bar x\right) +\xx\right); \nn \\
 \Delta\Phi_G\left(x,M^2\right) &=&
      \frac{\langle \alpha_s GG \rangle}{24\pi M^2}
       \left(\delta\left(x-\Delta\right)+ \xx \right), \\
 \Delta\varphi_{G}\left(x, M^2 \right) &=&
      \frac{\langle \alpha_s GG \rangle}{24\pi M^2}
 \left(\delta\left(\bar x\right)+\xx\right); \nn \\
 \Delta\Phi_G'\left(x,M^2\right) &=&
    \frac{\langle \alpha_s GG \rangle}{6\pi M^2}
     \frac{\theta\left(\Delta<x\right)\theta\left(x<1-\Delta\right)}
          {1-2\Delta};
\la{eq:phi_gs}\\
 \Delta\varphi_G'\left(x,M^2\right) &=&
    \frac{\langle \alpha_s GG \rangle}{6\pi M^2}. \nn
\ea
Here
$\Ds A_S=\frac{8\pi}{81}\langle\sqrt{\as}\bar q(0)q(0)\rangle^2$,
for quark and gluon condensates we use the standard estimates
$\langle\sqrt{\as}\bar q(0)q(0)\rangle\approx (-0.238\ \mbox{GeV})^3$,
$\Ds\frac{\langle\as GG\rangle}{12\pi}\approx 0.001$ GeV$^4$
\cite{svz} and
$\Ds\lambda_q^2
 =\frac{\langle\bar q\left(ig\sigma_{\mu\nu}G^{\mu\nu}\right)q\rangle}
         {2\langle\bar qq\rangle}
 =  0.4 \pm 0.1$~GeV$^2$, normalized at $\mu^2 \approx 1$~GeV$^2$.

\section{Expressions for perturbative spectral densities}
\la{subs-A.2}
 \setcounter{equation}{0}
First, $\rho(x,s)^{pert}_L$ in an order of $O(\as)$ was calculated in
\cite{nlwf,nlwf92},
but there was omitted the trivial term
$\Ds 2\ln\left[\frac{s}{\mu^2}\right]$ that follows from
the $\Ds \ln^2\left[\frac{-q^2}{\mu^2}\right]$-term
in correlators (\ref{cor-V})--(\ref{cor-T}); here we restore it.
The corresponding term for the $T$ case $\rho(x,s)^{pert}_T$
has recently been presented in \cite{bb96}.
We have recalculated it and confirmed this result.
\ba
 \rho_L^{Born}(x,s)
  &=& \frac{3}{2\pi^2} x\bar x, \\
 \rho_L^{pert}(x,s)
  &=& \frac{3}{2\pi^2} x\bar x
     \left\{1+\frac{\alpha_s(\mu^2)C_F}{4\pi}
      \left(2\ln\left[\frac{s}{\mu^2}\right]
            + 5 - \frac{\pi^2}{3}
            + \ln^2(\bar x/x)\right)\right\},
\la{eq:rcpi}                                \\
 \rho^{pert}_T(x,s)
  &=& \frac{3}{2\pi^2} x\bar x
     \left\{1+\frac{\alpha_s(\mu^2)C_F}{4\pi}
      \left(2\ln\left[\frac{s}{\mu^2}\right]
            + 6 - \frac{\pi^2}3 + \ln^2(\bar x/x) 
         +\ln(x\bar x)\right)\right\}.
\la{eq:rcrho}
\ea
Here $\mu^2 \sim 1$~GeV$^2$ corresponds to the average value of
the Borel parameter $M^2$ in the stability window;
$\alpha_s\left(1\mbox{GeV}^2\right)\approx 0.52$.

\end{appendix}

\newpage
\thispagestyle{empty}
\vspace{-117mm}\hspace{110mm}
 {\Large $\langle \xi^2 \rangle_{\rho}^L(M^2)$}
\newpage
\thispagestyle{empty}
\vspace{-118mm}\hspace{17mm}{\Large $m_{\rho'}^2$}
\newpage
\thispagestyle{empty}
\vspace{-110mm}\hspace{125mm}{\Large $\varphi_\rho^{L}(x)$}
\newpage
\thispagestyle{empty}
\vspace{-110mm}\hspace{25mm}{\Large $\varphi_{\rho'}^{L,mod}(x)$}
\end{document}